\documentclass{aa}
\bibpunct{(}{)}{;}{a}{}{,}
\usepackage[varg]{txfonts}
\usepackage{natbib,twoopt}
\usepackage{graphicx}   % Including figure files
\usepackage[dvipsnames]{xcolor}     % Extra text color
\usepackage{multirow}
\usepackage{placeins}
\newcommand{\Msun}[0]{\rm{M_\odot}}
\newcommand{\Lsun}[0]{\rm{L_\odot}}
\newcommand{\Rsun}[0]{\rm{R_\odot}}
\newcommand{\FeH}[0]{\rm{[Fe/H]}}
\newcommand{\MH}[0]{\rm{[M/H]}}

\newcommand{\Teff}[0]{T_\mathrm{eff}}

\begin{document}

\title{Studying the surface effect in Procyon A as an F-type star}
\author{Nuno Moedas\inst{1}\thanks{nuno.martinsmoedas@inaf.it}, \and Maria Pia Di Mauro\inst{1}}

\institute{INAF-IAPS, Via del Fosso del Cavaliere 100, 00133 Roma, Italy}

\date{Received XXX / Accepted YYY}

\abstract 
{Procyon A is an F-type main-sequence star in a binary system. It has been the subject of numerous ground-based and space-based observing campaigns, providing precise classical constraints, including a well-determined mass. It was also among the first stars in which individual frequencies were detected, making it a crucial benchmark for F-type stars.}
{Our goal is to investigate the surface effect, namely the discrepancy between observed and model oscillation frequencies due to inadequate modeling of the surface stellar layers, especially important in F-type stars. Using Procyon A as a case study, we aim to understand how different surface correction prescriptions impact the inference of the fundamental properties of this star, and compare the results with those obtained when the surface corrections are neglected.}
{We inferred the fundamental stellar properties employing a  grid of models computed with MESA, including gravitational settling, radiative accelerations, and turbulent mixing. We selected the best-fit models using the AIMS code taking into account different methods to fit the individual frequencies.}
{We find that the use of surface corrections can introduce uncertainties up to 7\% in the inferred stellar mass. We identify that the most reliable stellar mass estimates are obtained when using frequency ratios, the \cite{Sonoi2015} surface correction or directly fitting the individual frequencies.}
{Our results indicate that the surface effects in F-type stars differ from those found in the Sun and in solar-like stars, highlighting the need to be careful when considering the surface corrections for these stars.}
%Diffusion - Turbulence - Stars: abundances - Stars: evolution - Asteroseismology}

\titlerunning{Studying the surface effects in Procyon as an F-type star}
\authorrunning{Nuno Moedas \& Maria Pia Di Mauro}
\maketitle

\section{Introduction}
\label{sec:Intro}

Procyon is a binary star system consisting of an F-type main sequence primary and a  white dwarf secondary component. Procyon A has been the focus of various studies because it is the first star, other than the Sun, in which solar-like oscillations have been clearly detected \citep{Brown1991}.

Early attempts to resolve the seismic frequencies provided unsatisfactory outcomes.
Ground-based observations by \cite{Martic1999} succeeded to find the presence of oscillations, but were unable to resolve the individual modes. Successively, \cite{Eggenberger2004} confirmed  oscillations and identified few individual mode frequencies. Attempts to confirm the individual frequencies using space photometry were carried out with the Canadian MOST satellite \citep{Walker2003}; however, because of suboptimal noise levels and lower-than-expected oscillation amplitudes, it was not possible to resolve any excess of power \citep{Bedding2005}.
More recently, \citet{Bedding2010} analyzed the combined series of data collected from high-precision, ground-based observations, finally succeeding in identifying more than fifty individual oscillation frequencies. 
%The quality of these frequencies is comparable to that observed in the Sun. 
This accurate set of oscillation frequencies can be combined with a precise determination of the radius and luminosity from interferometry and parallax, as measured by the Very Large Telescope Interferometer \citep{Aufdenberg2005} and Gaia DR3 \citep{Soubiran2024}, together with spectroscopic determination of metallicity and effective temperature from HARPS \citep{Perdelwitz2024}. Furthermore, since Procyon is a binary system, the masses of both components are tightly constrained \citep{Girard2000,Gatewood2006,Bond2015}. Thanks to this wealth of high-quality data,
Procyon has become a unique testing ground for stellar models \citep[e.g.][]{Eggenberger2005,Straka2005,Guenther2014}. 

F-type stars are particularly challenging to study due to their distinct stellar structure, presenting a convective core during the main sequence (MS) and a very thin convective envelope. A major difficulty lies in the detection of solar-like oscillations, as their high effective temperatures lead to shorter mode lifetimes and broader line widths, which render the oscillation modes unresolved and hard to distinguish \citep{White2012}.

Only a limited number of F-type stars exhibit detectable individual oscillation mode frequencies. HD 49933, an F-type star similar to Procyon A, exhibits solar-like oscillations detected by the CoRoT mission \cite{Michel2008}. Along with Procyon, 
HD~49933 is among the most studied F-type stars, and has been widely used
as a laboratory for magnetic activity \citep[e.g.][]{Garcia2010,Ceillier2011}. 
Other F-type stars showing high-quality individual frequencies include 22 stars in the Kepler LEGACY sample \citep{Lund2017}, which represent the best seismic detections obtained with the \textit{Kepler} telescope \citep{kepler}.
These stars have been used in various studies to test stellar models and investigate different physical properties such as determining the depth of the convective zone \citep[e.g][]{Deal2023,Deal2025} and understanding the surface chemical evolution \cite[e.g.][]{Verma2019a,Verma2019,Moedas2025}. Nevertheless, Procyon A remains one of the few F-type stars with a dynamical mass determination. This is an invaluable constraint that can be used to validate stellar models.

With the upcoming launch of the PLAnetary Transits and Oscillations of stars space mission (PLATO/ESA; \citealt{Rauer2024}), high-precision asteroseismic data will become available, providing mass, radius, and age determinations with expected uncertainties of 15\% in mass, 2\% in radius, and 10\% in age for Sun-like stars, respectively. Achieving this precision requires accurate stellar models. 
However, it is important to note that current stellar models are still subject to significant uncertainty.
Therefore, it is necessary to identify and quantify the inaccuracies in stellar models and improve our modeling techniques.
One of the current challenges in stellar modeling is the treatment of surface effects. A well-known systematic discrepancy exists between the observed and modeled frequencies of the Sun and solar-type stars \citep{Christensen1988,Christensen1996,Dziembowski1988,Christensen1997}. This discrepancy is due to our inadequate modeling of the stellar surface layer. 
We still have an incomplete understanding of how to model the internal thermal gradient in the superadiabatic regions of stars. Currently, we use the adiabatic approximation to calculate the seismic frequencies. Additionally, we do not fully understand how to correctly model convection and the magnetic field in the surface layer. These issues are referred to as surface effects.
Several empirical formulas have been developed to correct for the discrepancies caused by these surface effects. See, for example,\cite{Kjeldsen2008,Ball2014,Sonoi2015}.
However, it should be noted that these surface corrections were developed for the Sun. While they may be applicable to solar-type stars, their validity for stars with different effective temperature may be limited, particularly in the case of F-type stars. These stars exhibit distinct stellar structures and possess a remarkably thin convective envelope, suggesting that the level and nature of surface effects may differ. Moreover, these corrections lack a direct physical interpretation.
In this study, we focus on Procyon A because its stellar parameters are well-known, especially its stellar mass. We use Procyon A to understand how surface corrections perform in inferring the global stellar properties and to determine whether the corrections are still valid for an F-type star.

This article is structured as follows: in Sect.~\ref{sec:Obsevable}, we present the most recent observable parameters of Procyon A. In Sect.~\ref{sec:Stellar_Models}, we describe the optimization method and the models used. The test we performed to select the frequencies scenarios is presented in Sect~\ref{sec:senarios}. Our results of testing the different surface corrections are presented in Sect.~\ref{sec:results}. Our conclusion is given in Sect.~\ref{sec:conclusion}.

\section{Observable Parameters}
\label{sec:Obsevable}

\begin{table}[t]
\centering
\caption{Procyon A fundamental parameters}
%\resizebox{\textwidth}{!}{%
\begin{tabular}{cc}
\hline
Parameter & Value \\ \hline
$M~(\Msun)$ & $1.478\pm0.012$ \\
$L~(\Lsun)$ & $7.049\pm0.064$ \\
$R~(\Rsun)$ & $2.046\pm0.009$ \\
$\Teff$~(K) & $6768.8\pm70.0$ \\
$\log(g)$~(dex) & $4.056\pm0.066$ \\
$\FeH$~(dex) & $0.00\pm0.056$ \\ \hline
\end{tabular}%
%}
\label{tab:Procyon_obs}
\end{table}

The Procyon system is a well-known binary star system. Thanks to astrometric measurements, the precise stellar masses of its two components can be determined. The most recent estimates, based on Hubble Space Telescope data, were provided by \citet{Bond2015}, who derived masses of $1.478\pm0.012~\Msun$ for Procyon A and $0.592\pm0.006~\Msun$ for Procyon B. These values provide strict constraints for validating stellar evolution and asteroseismic models.

\begin{figure}
    \centering
    \includegraphics[width=\columnwidth]{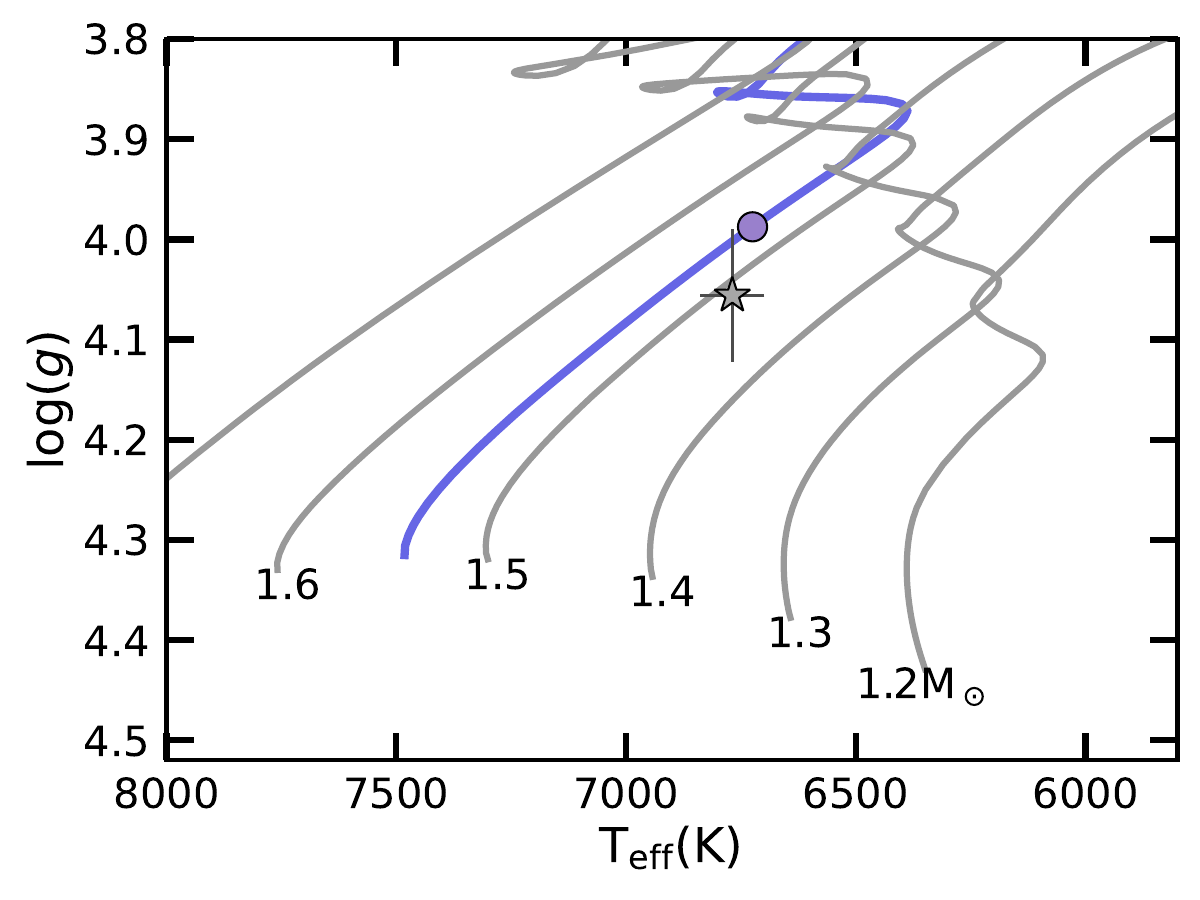}
    \caption{Kiel diagram showing some computed evolutionary tracks with metallicity $\MH_i=0.0$ and $Y_i=0.26$ (solar composition) in solid gray lines. The star symbol shows the Procyon A's location in the diagram. The blue solid line is the evolutionary track of the best model (circle symbol) from the $r_{02}$ test considering radius in the inference of Scenario~A.}
    \label{fig:Kiel_Procyon}
\end{figure}

To constrain the primary component, Procyon A, we adopted the most recent spectroscopic and photometric measurements.
The stellar luminosity $(L=7.049\pm0.064~\Lsun)$ and the radius ($R=2.046\pm0.009\Rsun$), were determined by \cite{Soubiran2024} using Gaia parallaxes and interferometric angular diameters \cite{Gaia_Collaboration2021}. Spectroscopic parameters such as effective temperature ($\Teff$), surface gravity ($\log(g)$), and iron content ($\FeH$), were taken from \cite{Perdelwitz2024} who report $\Teff=6768.8\pm70.0$~K, $\log(g)=4.056\pm0.066$~dex, and $\FeH=0.00\pm0.056$~dex (see Table \ref{tab:Procyon_obs}). Figure \ref{fig:Kiel_Procyon} shows the location of Procyon A in the Kiel diagram.

High-precision asteroseismic observations by \citet{Bedding2010} revealed a set of individual p-mode frequencies $(\nu_{n,\ell})$  with radial order $n$ and harmonic degree $\ell$ in Procyon A, based on combined observed data from 11 telescopes \citep{Arentoft2008}. Two alternative mode identifications were proposed, referred to as Scenario A and Scenario B, each including approximately fifty modes of degree $\ell = 0$, 1, and 2. Due to challenges in ridge identification in the \'echelle diagram, a plot in which oscillation frequencies are folded modulo the large frequency separation to reveal regular patterns of modes, both scenarios were considered, focusing on the $\ell = 0$ and $\ell = 1$ modes. From an observational point of view, \citet{Bedding2010} and \citet{White2012} found some evidence favoring Scenario B. However, several modeling studies \citep{Dogan2010, Guenther2014, Compton2019} have shown that only Scenario A leads to acceptable stellar models that are consistent with the observed global and seismic parameters.

\section{Stellar models and inference method}
\label{sec:Stellar_Models}

\subsection{Stellar models}
In this manuscript, we use grid C of the stellar models developed in \cite{Moedas2025}, which was computed using the Modules for Experiments in Stellar Astrophysics (MESA) code \citep{Paxton2011,Paxton2013,Paxton2015,Paxton2018,Paxton2019}. 
This grid considers up-to-date stellar physics. In particular, it includes atomic diffusion and the effects of radiative acceleration, which are calculated using the Single-Valued Parameter (SVP) method \citep{LeBlanc2004, alecian20}. To prevent unrealistic surface chemical variations caused by diffusion, the grid uses a turbulent mixing prescription calibrated by \citet{Verma2019} to reproduce the observed surface helium abundance in F-type stars.

The impact of using chemical transport mechanisms, such as radiative accelerations and turbulent mixing, to characterize F-type stars was investigated in \cite{Moedas2024,Moedas2025}.  Using a sample of 91 FGK-type stars from \cite{Davies2016,Lund2017}, the authors found that neglecting atomic diffusion in stellar models of F-type stars could introduce uncertainties of up to 10\%, 4\%, and 29\% when inferring the stellar mass, radius, and age, respectively. Since this is outside the scope of our work, we do not investigate how considering chemical transport mechanisms affects modeling Procyon A; however, the results and conclusions should be consistent when the same physical inputs are used in stellar models.

\subsection{Optimization}

For the inference of the stellar properties, we employed the AIMS code \citep{Rendle2019} following the approach of \citet{Moedas2024,Moedas2025}. Here we give a brief overview of the optimization process. AIMS is a Bayesian optimization tool that uses  Markov chain Monte Carlo (MCMC) methods to explore the parameter space of stellar model grids and identify the model that best fits the observational constraints.
AIMS distinguishes the contribution of the global constraints, $X_i$ (in our case $\Teff$, $\FeH,$ and $R$),
\begin{equation}
    \label{eq:Chi_class} 
    \chi^2_\mathrm{global}=\sum^3_i\left(\frac{X_i^\mathrm{(obs)}-X_i^\mathrm{(mod)}}{\sigma(X_i)}\right)^2
\end{equation}
and the constraints from individual frequencies, $\nu_{i}$,
\begin{equation}
    \label{eq:Chi_seis} 
    \chi^2_\mathrm{freq}=\sum^N_i\left(\frac{\nu_{i}^\mathrm{(obs)}-\nu_{i}^\mathrm{(mod)}}{\sigma(\nu_{i})}\right)^2,
\end{equation}
where "(obs)" corresponds to the observed values and (mod) corresponds to the calculated model values.
The weight that AIMS gives to the seismic contribution can be absolute (3:N), where each individual frequency has the same weight as each global constraint,
\begin{equation}
    \label{eq:Chi_tot_abs} 
    \chi^2_\mathrm{total}=\chi^2_\mathrm{freq}+\chi^2_\mathrm{global}.
\end{equation}

% NEW: Clarification on raw chi-square requested by referee (Point 2 and 4)

It is important to note that the $\chi^2$ values reported in this work (and provided by AIMS) are the raw sum of squared residuals, not divided by the number of degrees of freedom (reduced $\chi^2$). Therefore, they should be interpreted primarily as the objective function minimized by the Bayesian optimization engine to select the best-fit models, rather than as a standard goodness-of-fit statistic for strictly comparing models with different degrees of freedom.

To interpret the statistics, we can calculate a reduced $\chi^2_\mathrm{r}$
\begin{equation}
    \label{eq:Chi_tot_rel} 
    \chi^2_\mathrm{r}=\frac{\chi^2}{N},
\end{equation}
where $N$ is the number of constraints in the calculation of the $\chi^2$.

\subsubsection{Data analysis}

In this study, since we are using a single star to compare surface corrections, we cannot make strong statistical claims. 
%However, since we know some stellar properties of Procyon (such as its mass, which was determined by interferometry), we can evaluate how well each test characterizes this star. }
However, given that several stellar properties of Procyon A, such as its mass determined from interferometry, are well known, we can evaluate how accurately each surface prescription characterizes this star.
To compare the inferred results with observational parameters, we computed the relative difference, defined as:
\begin{equation}
    \frac{\Delta X}{X_r}=\frac{X_m - X_r}{X_r},
\end{equation}
where $X$ represents a given stellar parameter, $X_m$ is the value inferred from the model, and $X_r$ is the reference observational value. This provides a measure of the discrepancy between the model prediction
and the
reference observed values. Additionally, we can implement the use of a normalized difference:
\begin{equation}
    z=\frac{|X_m - X_r|}{\sigma(X_m)},
    \label{eq:norm}
\end{equation}
where $\sigma(X_m)$ is the uncertainty of the inferred parameter. This quantity indicates how the model result deviates from the expected values in units of the 1$\sigma$ uncertainty. 
%However, this parameter is sensitive to the internal uncertainties of the model. Therefore, we use it only as a qualitative indicator.}
We caution, however, that $z$ relies on the model-inferred uncertainties $\sigma(X_m)$, which can vary significantly between different models and parameters depending on the shape of the posterior distribution. Therefore, $z$ is not a standard statistical metric and should not be used as a primary diagnostic for model selection; rather, we use it as a qualitative indicator to easily visualize the agreement between the inferred and reference values.
\subsection{Surface Corrections}

A well-known issue in asteroseismic modeling is the surface effect, which is a systematic discrepancy ($\delta\nu$) between observed and theoretical frequencies of individual stars arising from simplified modeling of the near-surface stellar layers.
%due to the simplification of near-surface layers in the stellar models.
Two main approaches are used to compare observations and models. The first involves applying empirical surface corrections to compensate for the offset between modeled and observed frequencies, while the second relies on comparing ratios of small to large frequency separations rather than the individual frequencies.
Several surface correction prescriptions exist; in this work, we investigate the three most commonly applied formulations.

The first correction we consider was developed by \citet[hereafter KJ08]{Kjeldsen2008}. They found that, using the Sun as a reference, the offset between the observed and model frequencies could be described by a power law:
\begin{equation}
    \delta\nu=a\left(\frac{\nu_i^\mathrm{(obs)}}{\nu_\mathrm{max}} \right)^b,
\end{equation}
where $a$ and $b$ are the correction parameters, $\nu_\mathrm{(max)}$ is the maximum power frequency.

The second prescription was proposed by \citet[hereafter BG14]{Ball2014}. In their approach, the discrepancy was described by two terms that take into account the mode inertia ($I$) of the mode. The correction is described as follows:
\begin{equation}
\label{eq:BG142}
\delta\nu=\left[a_3\left(\frac{\nu_i}{\nu_\mathrm{ac}}\right)^3+a_{-1}\left(\frac{\nu_i}{\nu_\mathrm{ac}}\right)^{-1}\right]/I,
\end{equation}
where $a_3$ and $a_{-1}$ are the correction parameters and $\nu_\mathrm{ac}$ is the acoustic cut-off frequency,  which is the maximum frequency at which pressure modes can be reflected back into the stellar interior rather than escaping into the atmosphere.
The first term describes the contribution of the magnetic field ($\propto\nu_i^3$) and the second term accounts for the contribution of the convection ($\propto\nu_i^{-1}$). BG14 showed that the first term dominates the correction, allowing the second term to be neglected and the Eq. (\ref{eq:BG142}) to be simplified as follows:
\begin{equation}
\label{eq:BG141}
\delta\nu=\left[a_3\left(\frac{\nu_i}{\nu_\mathrm{ac}}\right)^3\right]/I.
\end{equation}
Hereafter, we shall refer to the two-term correction (Eq. \ref{eq:BG142}) as  BG14$_2$ and the one-term correction (Eq. \ref{eq:BG141}) as BG14$_1$.

The final prescription we investigate was proposed by \citet[hereafter SO15]{Sonoi2015}. SO15 pointed out that previously formulated surface corrections depend strongly on the star's $\Teff$ and $\log(g)$, since the structure of the stellar surface varies with these parameters. They also emphasized that the surface corrections should not be calibrated on the Sun, but rather constrained by model physics. 
Based on 3D models of stellar surface layers, SO15 proposed a modified Lorentzian function to describe the frequency offsets:
\begin{equation}
    \frac{\delta\nu}{\nu_\mathrm{max}}=\alpha\left[1-\frac{1}{1-\left(\nu_\mathrm{(obs)}/\nu_\mathrm{max} \right)^\beta}\right]
\end{equation} 
where $\alpha$ and $\beta$ are free parameters. %It is important to mention that all the prescriptions could perform well in the fitting process however find a minimum in the optimization that far from the real results.

The second method for comparing stellar models and asteroseismic observations was introduced by \cite{Roxburgh2003} and is based on the ratio of the small to the large frequency separations. The most commonly used small separation is calculated as the difference between two modes with consecutive radial order $n$ but different $\ell$:
\begin{equation}
    \label{eq:smal_sep}
    d_{02}(n)=\nu_{n,0}-\nu_{n-1,2},
\end{equation}
where $\nu_{n,0}$ is the frequency of the radial mode ($\ell=0$) of radial order $n$, and $\nu_{n-1,2}$ is the frequency of the quadrupole mode ($\ell=2$) of radial order $n-1$. 

This quantity is very useful because it is sensitive to the structure of the stellar core, being defined by the integral of the sound-speed gradient in the core, which depends on  the composition gradients and hence directly related to the evolutionary stage of the star:
\begin{equation}
\mathrm{d}_{02}(n) \propto \int_0^R \frac{dc(r)}{dr} \frac{dr}{r},
\end{equation}
where $c(r)$ is the sound speed and $r$ the radius.
In order to be calculated Eq.~(\ref{eq:smal_sep}) requires the detection of modes with $\ell=2$, which are not always observable. 
Hence, other small separations can be considered, such as $d_{01}$ and $d_{10}$, which are defined by the so-called five-point separations:
\begin{equation}    
\label{eq:5_p_fre}
    \begin{gathered} 
d_{01}(n)=\frac{1}{8}(\nu_{n-1,0}-4\nu_{n-1,1}+6\nu_{n,0}-4\nu_{n,1}+\nu_{n+1,0}),\\
d_{10}(n)=-\frac{1}{8}(\nu_{n-1,1}-4\nu_{n-1,0}+6\nu_{n,1}-4\nu_{n,0}+\nu_{n+1,1}).
    \end{gathered}
\end{equation}
%\eq{\label{eq:5_p_fre}}
With the small separations we can define the frequency ratios as
\begin{equation}
    \label{eq:ratios}
    r_{01}(n)=\frac{d_{01}(n)}{\Delta \nu_1(n)},\hspace{7pt}
    r_{10}(n)=\frac{d_{10}(n)}{\Delta \nu_0(n+1)},\hspace{7pt}
    r_{02}(n)=\frac{d_{02}(n)}{\Delta \nu_1(n)},
\end{equation}
where $\Delta\nu_\ell(n)$ is the large frequency separation, defined by the asymptotic relation \citep{Tassoul1980}:
\begin{equation}
    \Delta\nu_\ell(n)\propto\left(2 \int_0^R \frac{dr}{c}\right)^{-1}\approx\nu_{n,\ell}-\nu_{n-1,\ell}.
    \label{eq:large}
\end{equation}
For a more detailed description, see \cite{Roxburgh2003}. These ratios are a set of constraints that are more sensitive to the interior of the star and nearly independent of the outer layer structure. They allow us to avoid surface corrections and compare models to observations more robustly. However, a sufficiently large set of individual frequencies is required, and the ratios lose some information about the outer layers, making it more difficult to accurately infer the mean density of the observed star. %It is important to note that the chi-squared statistic ($\chi^2_\mathrm{freq}$) is not calculated from the individual frequencies, but rather from the computed ratios.

\begin{figure}
    \centering
    %\fbox{}
    \includegraphics[width=1\linewidth]{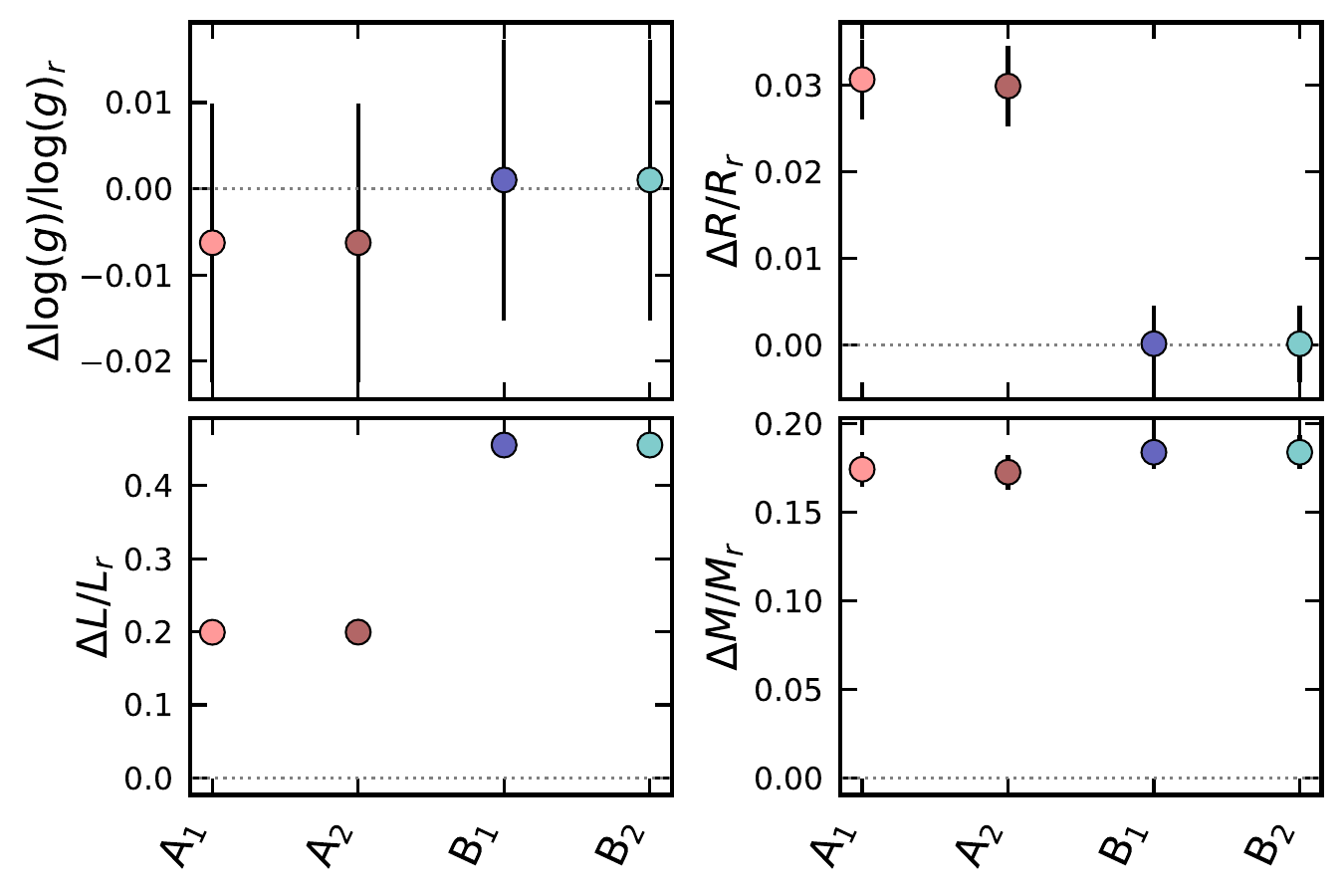}
    \caption{Relative difference between the inferred and reference values (subscript $r$) obtained in test~1 for the two sets of frequencies (Scenario A and B). Top left panel shows the surface gravity, top right panel the surface radius, bottom left panel the surface luminosity, and bottom right panel the stellar mass. The dotted line indicate where the relative difference is 0.}
    \label{fig:diff_set_all}
\end{figure}
\begin{figure}
    \centering
    \includegraphics[width=1\linewidth]{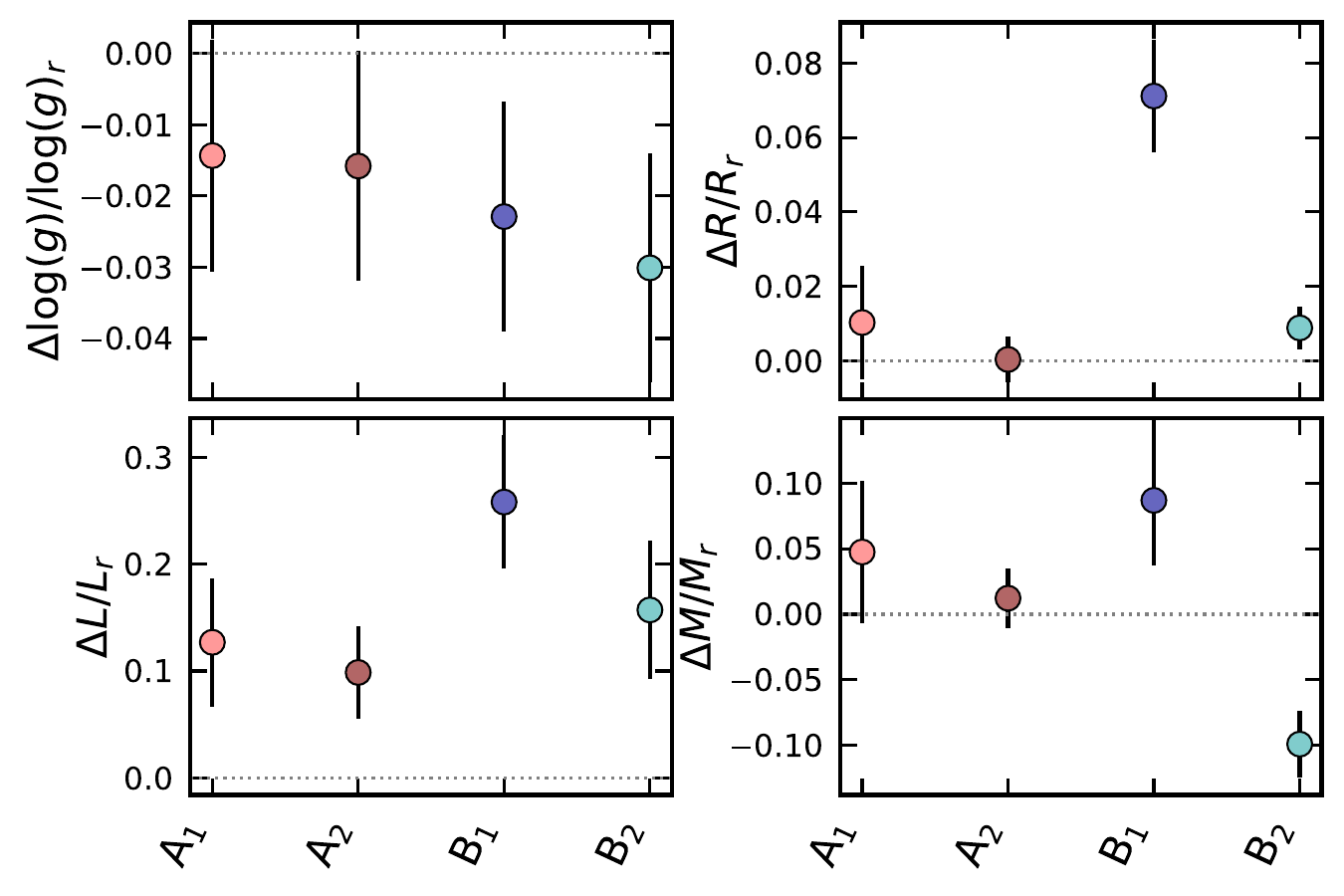}
    \caption{Same plots as Fig. \ref{fig:diff_set_all}, but showing the relative differences obtained in test 2.}
    \label{fig:diff_set_l02}
\end{figure}
\begin{figure}
    \centering
    \includegraphics[width=1\linewidth]{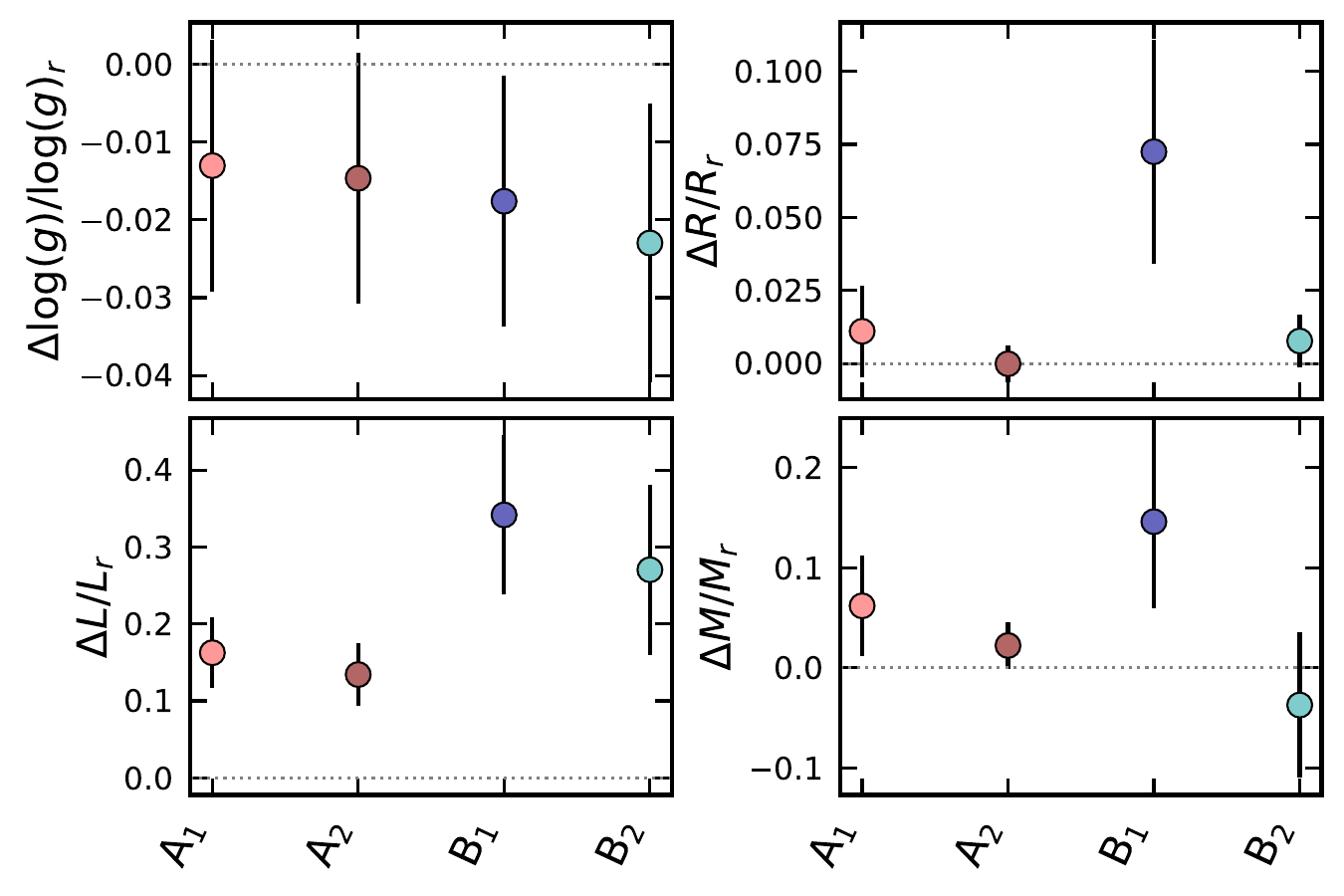}
    \caption{Same plots as Fig. \ref{fig:diff_set_all}, but showing the relative differences obtained in test 3.}
    \label{fig:diff_set_Al1_Bl02_red}
\end{figure}

\section{Scenario A vs scenario B}
\label{sec:senarios}
Our goal is to evaluate how different surface corrections perform in inferring stellar properties. However, we must adopt one of the frequency sets provided in \cite{Bedding2010}. To avoid dependence on surface corrections, we tested both frequency scenarios using frequency ratios, which are independent of surface corrections and thus provide more robust constraints on the stellar interior.
Here, we perform three tests.

In Test 1 the three frequency ratio ($r_{01}$, $r_{10}$, $r_{02}$) were used, and therefore all the mode frequencies were included. In Test 2 the $r_{02}$ was used and only modes with $\ell=0$ and $\ell=2$ were considered, excluding $\ell=1$ modes in the optimization process. This avoids issues arising from mixed modes or wrong mode identification. 
In Test 3 ($r_{\ell,red}$), we neglected the frequencies above which a  discontinuity in the solar-like asymptotic behavior occurs (Eq. \ref{eq:large}).
%the high-frequency modes, where there is a discontinuity in their asymptotic behavior, of the $\ell=1$ modes for Scenario A and the $\ell=0$ and $\ell=2$ modes for Scenario B.
%We will explain the purpose of each test as we present them.
In order to infer the stellar properties, the frequency ratios are used as seismic constraints. These constraints are computed from the individual mode frequencies corresponding to each scenario. For the classic constraints, we performed two tests: one considering $\Teff$ and $\FeH$, and another that included $R$ as an additional constraint.
For each test, we obtained four sets of inferred stellar properties: two for each scenario depending on whether $R$ was included as a constraint. This setup allows us to investigate the impact of including $R$  on the results. The results of each test are shown in Table~\ref{tab:set_results}, while the relative and normalized differences are reported in Table~\ref{tab:test_diffs}.

\begin{table*}[]
\centering
\caption{The inferred stellar properties from each test for the two frequency scenarios.  The index 1 is respect to the test considering the $\Teff$ and $\FeH$ in the optimization and index 2 respect to the test that we also considered $R$ in the optimization process. The $\chi_\mathrm{total}^2 $ and $\chi_\mathrm{freq}^2$ are from the best model provided by AIMS and there respective reduced value}.
\label{tab:set_results}
\begin{tabular}{cccccccccc}
\hline
Test & Scenario & $M~(\Msun)$ &  $R~(\Rsun)$ & $\log(g)$~(dex)& $L~(\Lsun)$& $\chi_\mathrm{total}^2 $& $\chi_\mathrm{freq}^2$& $\chi_\mathrm{total;r}^2 $& $\chi_\mathrm{freq;r}^2$ \\ \hline
\multirow{4}{*}{$r_{01}$, $r_{10}$, $r_{02}$} & A$_1$ & $1.736\pm0.002$ & $2.106\pm0.002$ & $4.031\pm0.001$ & $8.45\pm0.01$ & 818 & 756 & 18.2 & 17.6\\
 & A$_2$ & $1.733\pm0.004$ & $2.104\pm0.002$ & $4.031\pm0.001$ & $8.46\pm0.01$ & 864 & 756 & 18.9 & 17.6\\ \cline{2-10} 
 & B$_1$ & $1.750\pm0.001$ & $2.043\pm0.001$ & $4.060\pm0.001$ & $10.26\pm0.01$ & 1560 & 1490 &34.7 &34.7\\
 & B$_2$ & $1.750\pm0.001$ & $2.043\pm0.001$ & $4.060\pm0.001$ & $10.26\pm0.01$ & 1599 & 1490 &35.5 &34.7\\ \hline
\multirow{4}{*}{$r_\mathrm{02}$} & A$_1$ & $1.548\pm0.079$ & $2.064\pm0.030$ & $3.998\pm0.011$ & $7.94\pm0.41$ & 12 & 11 &0.6&0.6\\
 & A$_2$ & $1.496\pm0.033$ & $2.044\pm0.009$ & $3.992\pm0.007$ & $7.74\pm0.28$ & 15 & 14&0.7&0.8\\ \cline{2-10} 
 & B$_1$ & $1.606\pm0.127$ & $2.188\pm0.029$ & $3.963\pm0.010$ & $8.87\pm0.43$ & 17 & 16&0.9&0.9\\
 & B$_2$& $1.332\pm0.036$ & $2.061\pm0.007$ & $3.934\pm0.011$ & $8.16\pm0.45$ & 33 & 25&1.6&1.5\\ \hline
\multirow{4}{*}{$r_\mathrm{\ell,red}$} & A$_1$ & $1.570\pm0.074$ & $2.066\pm0.031$ & $4.003\pm0.009$ & $8.81\pm0.31$ & 18 & 17&0.6& 0.6\\
 & A$_2$ & $1.511\pm0.033$ & $2.043\pm0.009$ & $3.997\pm0.007$ & $7.99\pm0.28$ & 21 & 20& 0.7& 0.7\\ \cline{2-10} 
 & B$_1$ & $1.693\pm0.127$ & $2.191\pm0.078$ & $3.985\pm0.010$ & $9.46\pm0.72$ & 157 & 145 &5.8&5.8\\
 & B$_2$ & $1.423\pm0.107$ & $2.059\pm0.016$ & $3.963\pm0.011$ & $8.95\pm0.78$ & 180 & 134  &6.4&5.4\\ \hline
\end{tabular}
\end{table*}

For Test 1, the relative differences are shown in Fig. \ref{fig:diff_set_all}. When using all the frequency ratios, the model fails to reproduce the stellar properties accurately (except for $\log g$), obtaining unrealistic values for the stellar mass. Including the radius as an additional constraint in the fitting procedure does not improve the stellar inference, as all tests show a relative mass difference of about 17-18\%. Only Scenario B is able to recover the stellar radius, even without explicitly using it as a constraint.  Moreover, the normalized differences show that none of the inferred parameters (except $R$ in Scenario B) fall within the expected observational uncertainties, either at the $1\sigma$ ($z \leq 1$) or $3\sigma$ ($z \leq 3$) level, with $z$ defined in eq.\ref{eq:norm}. Nevertheless, the $\chi^2$ values for both frequency sets are extremely large, suggesting poor convergence in the optimization process and that the difference in frequencies is the main contributor to the $\chi^2$ estimate. This may point to an issue with the seismic constraints themselves. As noticed by \cite{Bedding2010}, the presence of a low-frequency mixed mode for $\ell=1$  could introduce inconsistencies during the optimization process.

\begin{table*}[]
\centering
\caption{Relative and normalized differences of the inferred stellar properties for each test and for the two frequency scenarios. Index 1 refers to the test considering $\Teff$ and $\FeH$ in the optimization and index 2 refers to the test that also includes $R$ in the optimization process.}
\label{tab:test_diffs}
%\resizebox{\textwidth}{!}{%
\begin{tabular}{cccccccccc}
\hline
\multirow{2}{*}{Test} & \multirow{2}{*}{Scenario} & \multicolumn{2}{c}{Mass} & \multicolumn{2}{c}{Radius} & \multicolumn{2}{c}{$\log(g)$} & \multicolumn{2}{c}{Luminosaty} \\ \cline{3-10} 
 &  & $\Delta M/M_r$ & $z$ & $\Delta R/R_r$ & $z$ & $\Delta \log(g)/\log(g)_r$ & $z$ & $\Delta L/L_r$ & $z$ \\ \hline
\multirow{4}{*}{$r_{01}$, $r_{10}$, $r_{02}$} & A$_1$ & $0.17\pm0.01$ & 96 & $0.031\pm0.005$ & 42 & $-0.006\pm0.016$ & 63 & $0.20\pm0.01$ & 128 \\
 & A$_2$ & $0.17\pm0.01$ & 70 & $0.030\pm0.005$ & 34 & $-0.006\pm0.016$ & 70 & $0.20\pm0.01$ & 102 \\
 & B$_1$ & $0.18\pm0.01$ & 1087 & $0.000\pm0.004$ & 0.6 & $0.001\pm0.016$ & 15 & $0.45\pm0.01$ & 226 \\
 & B$_2$ & $0.18\pm0.01$ & 1087 & $0.000\pm0.004$ & 0.6 & $0.001\pm0.016$ & 15 & $0.45\pm0.01$ & 223 \\ \hline
\multirow{4}{*}{$r_\mathrm{02}$} & A$_1$ & $0.04\pm0.05$ & 0.89 & $0.010\pm0.010$ & 0.70 & $-0.014\pm0.016$ & 5.07 & $0.13\pm0.06$ & 2.14 \\
 & A$_2$ & $0.01\pm0.02$ & 0.58 & $0.000\pm0.006$ & 0.09 & $-0.016\pm0.016$ & 8.87 & $0.10\pm0.04$ & 2.35 \\
 & B$_1$ & $0.09\pm0.05$ & 1.77 & $0.071\pm0.015$ & 4.93 & $-0.023\pm0.016$ & 9.19 & $0.26\pm0.06$ & 4.20 \\
 & B$_2$ & $-0.10\pm0.03$ & 4.02 & $0.009\pm0.006$ & 2.46 & $-0.030\pm0.016$ & 10.77 & $0.16\pm0.06$ & 2.46 \\ \hline
\multirow{4}{*}{$r_\mathrm{\ell,red}$} & A$_1$ & $0.06\pm0.05$ & 1.24 & $0.011\pm0.016$ & 0.7 & $-0.012\pm0.016$ & 5.74 & $0.16\pm0.05$ & 3.64 \\
 & A$_2$ & $0.02\pm0.02$ & 1.01 & $-0.000\pm0.006$ & 0.03 & $-0.014\pm0.016$ & 8.25 & $0.13\pm0.04$ & 3.41 \\
 & B$_1$ & $0.14\pm0.09$ & 1.70 & $0.072\pm0.076$ & 1.9 & $-0.017\pm0.016$ & 7.42 & $0.34\pm0.10$ & 3.32 \\
 & B$_2$ & $-0.04\pm0.07$ & 0.5 & $0.007\pm0.009$ & 0.97 & $-0.022\pm0.018$ & 2.80 & $0.27\pm0.11$ & 2.46 \\ \hline
\end{tabular}%
%}
\end{table*}

\begin{figure}
    \centering
    \includegraphics[width=\linewidth]{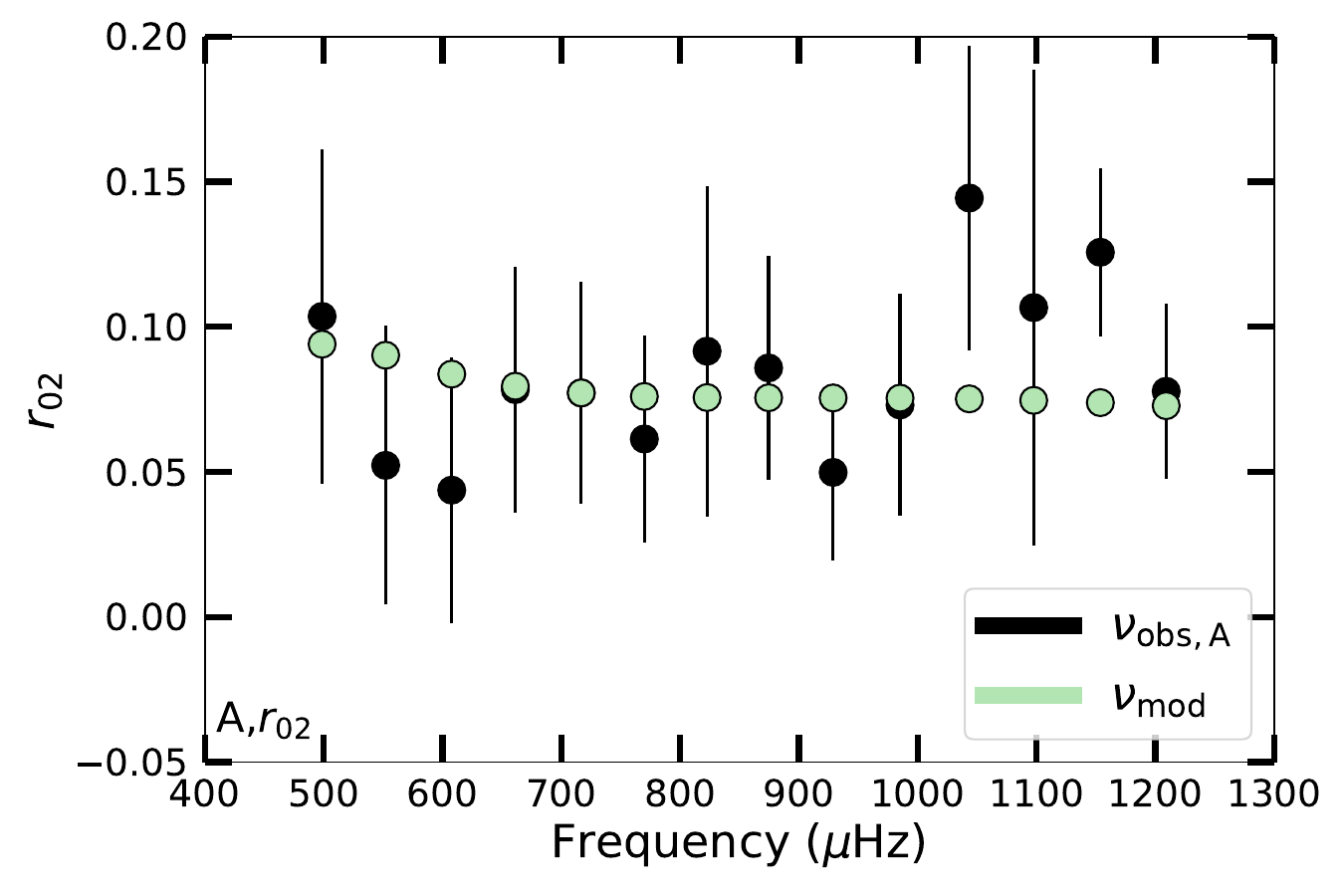}
    \includegraphics[width=\linewidth]{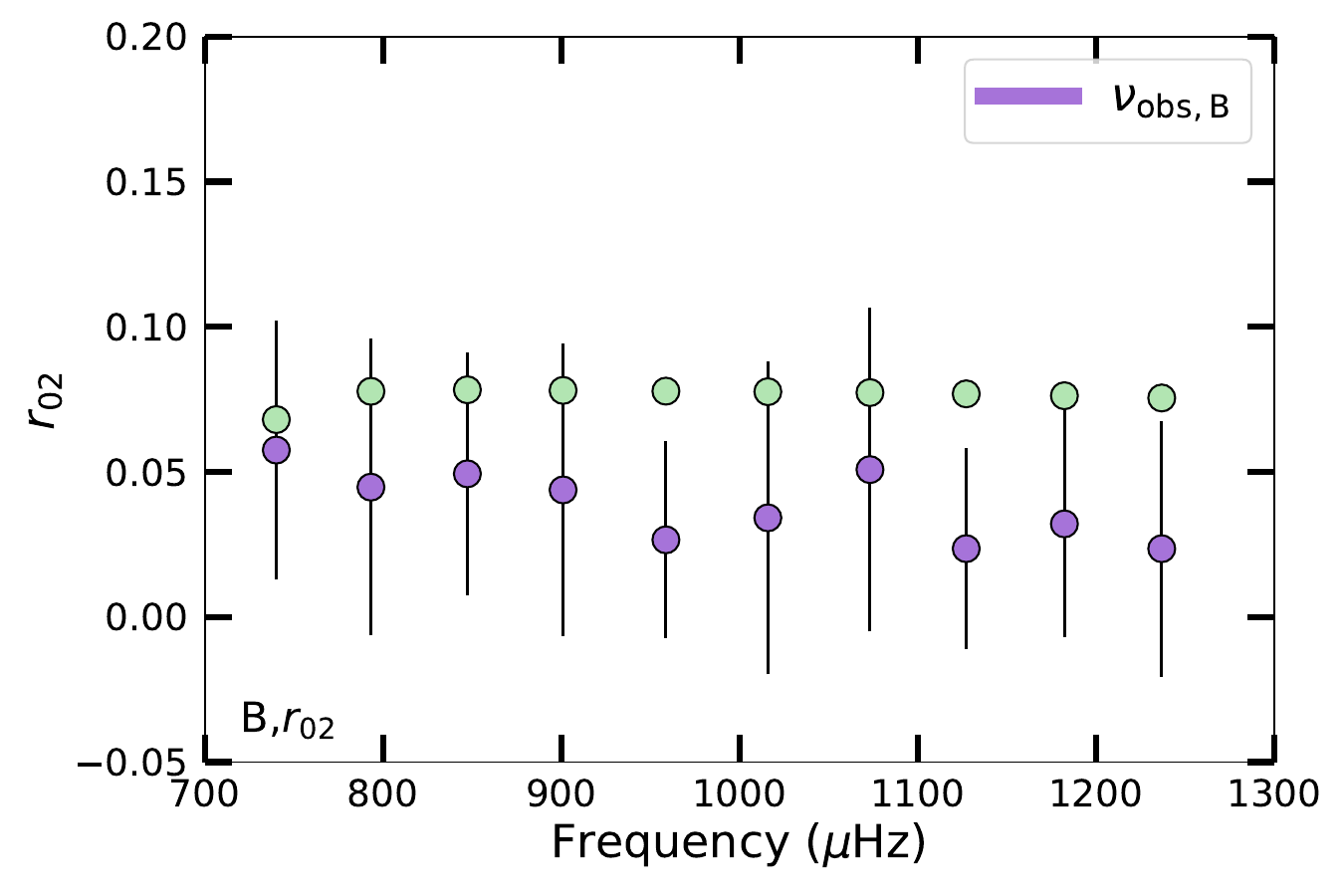}
    \caption{$r_{02}$ ratios of test 2 for Scenario A (top panel) and Scenario B (bottom panel). The black points with error bars are computed from the observed frequencies and green points for the best-fit model.}
    \label{fig:ratios_fit}
\end{figure}

\begin{figure*}
    \centering
    \includegraphics[width=.44\linewidth]{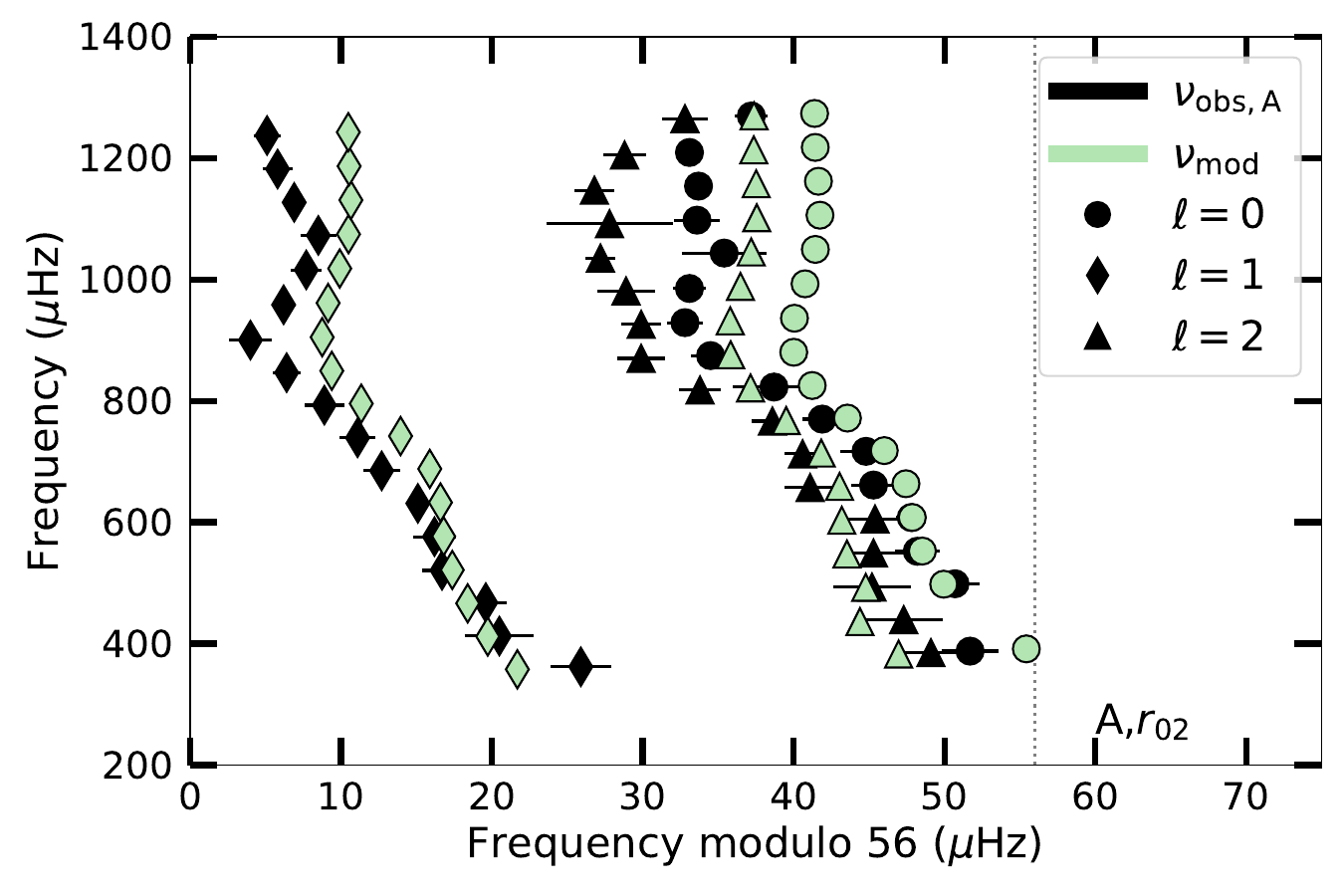}
    \includegraphics[width=.44\linewidth]{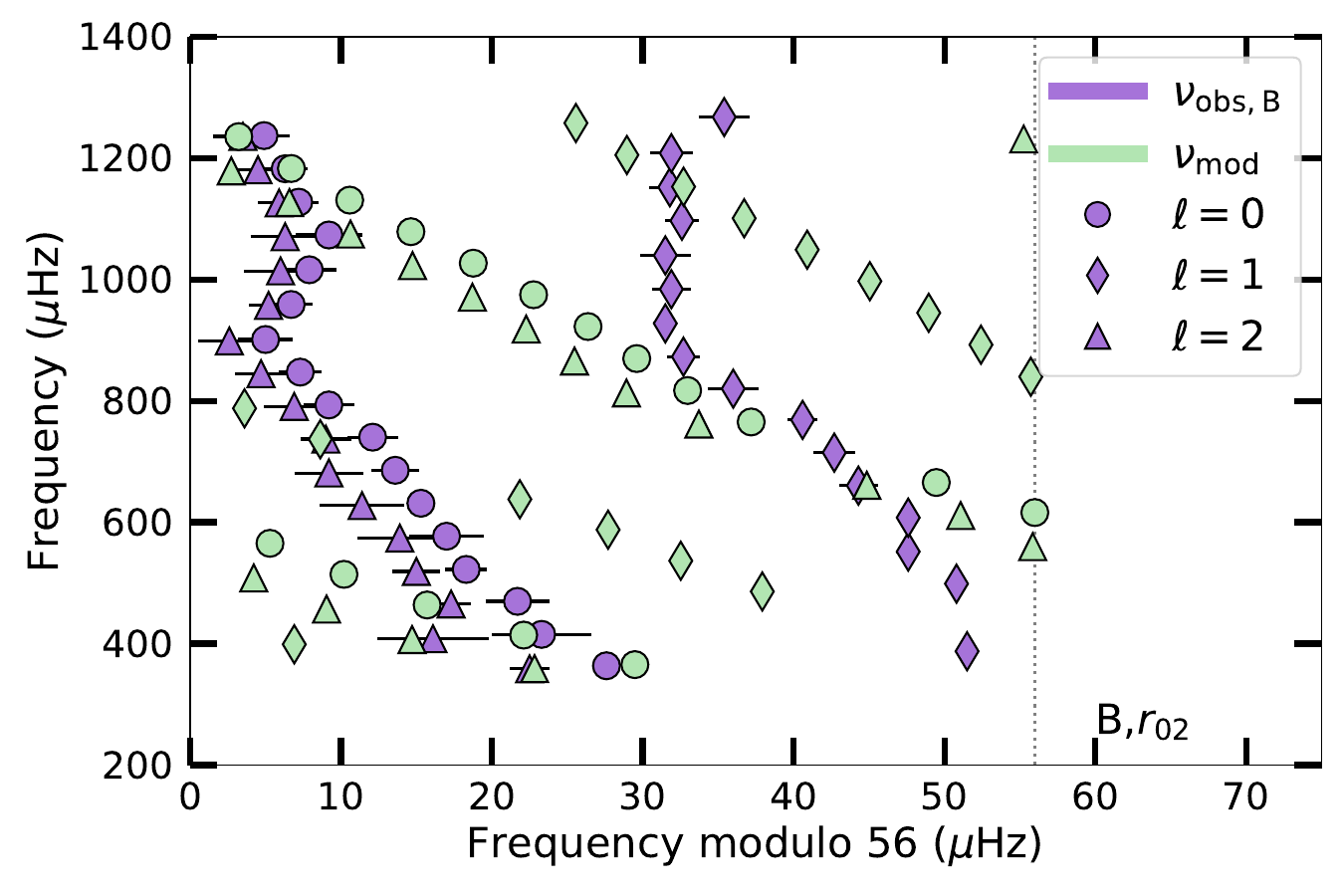}
    \includegraphics[width=.44\linewidth]{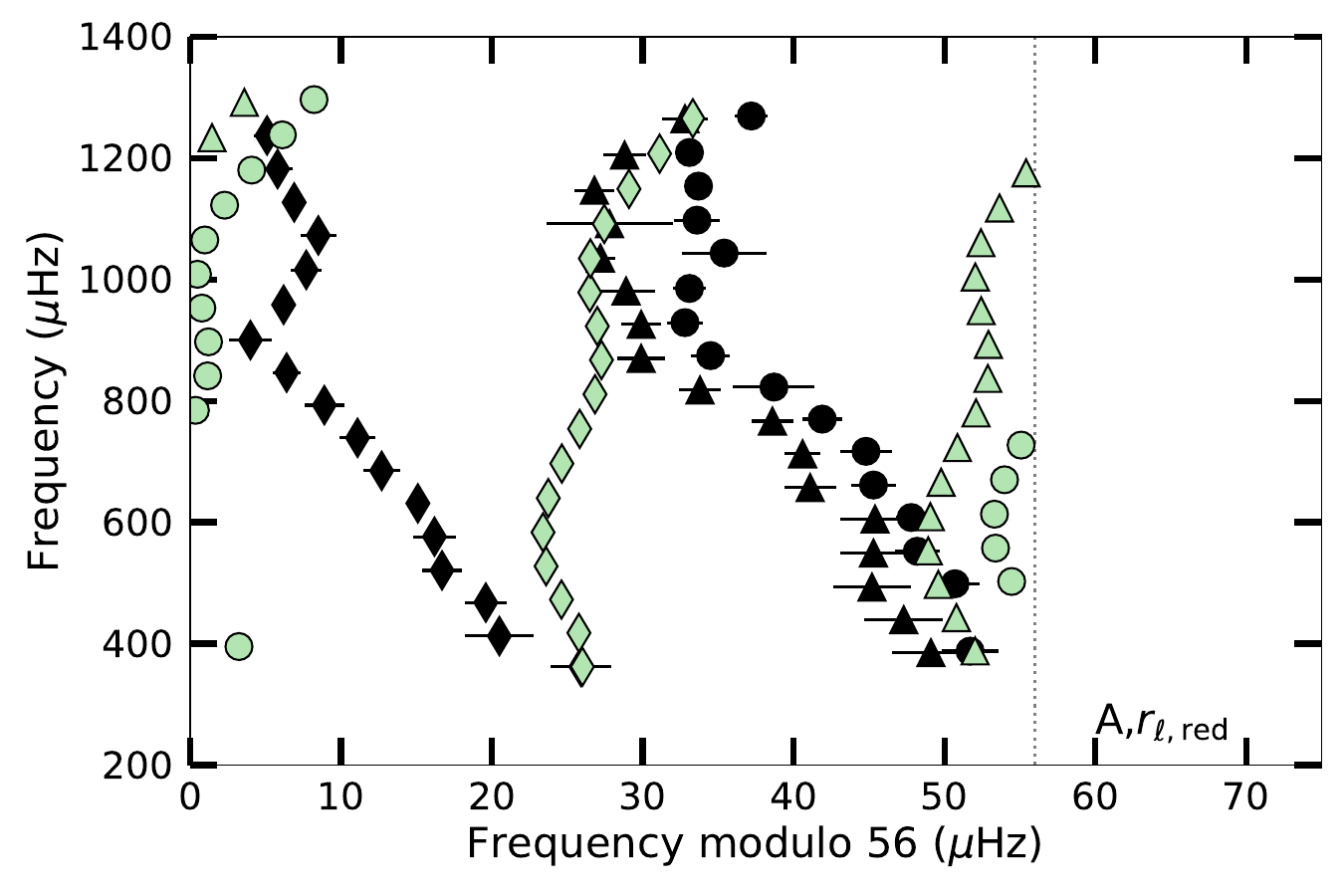}
    \includegraphics[width=.44\linewidth]{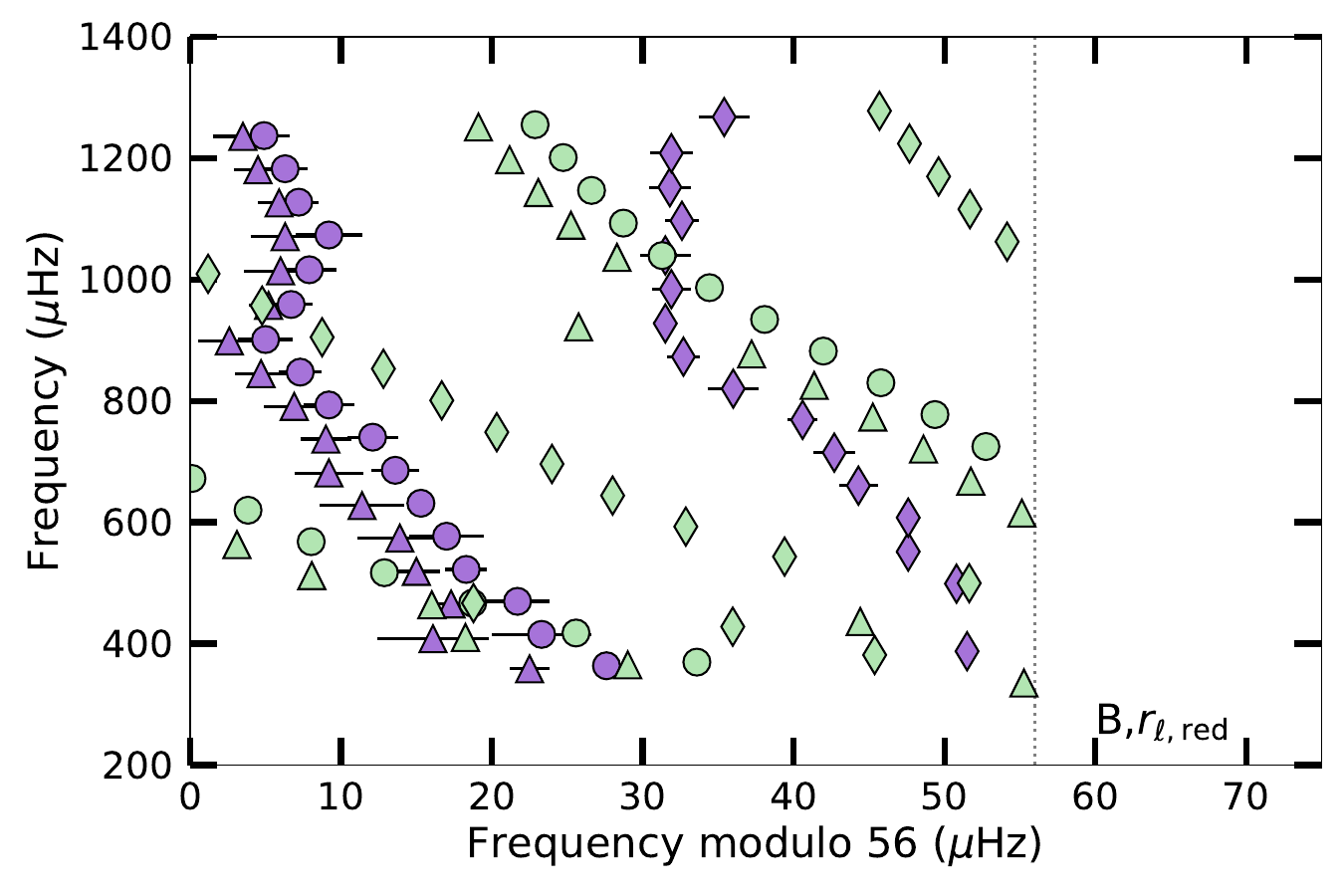}
    \caption{\'Echelle diagram of the frequencies of the best models from test 2 (top row) and 3 (bottom row) with the ratios. Left columns are frequencies from Scenario A and right are the frequencies from Scenario B. The circle represent modes $\ell=0$, triangles represent modes $\ell=1$ and diamond represent modes $\ell=2$. The black symbols with error bar are observed frequencies, green symbols represent the model frequencies. The gray dotted line indicates the 56~$\mu$HZ frequency.}
    \label{fig:echelle_set_test}
\end{figure*}

Figure \ref{fig:diff_set_l02} shows the results for Test 2, where we observe significant improvements in the determination of the stellar parameters. For Scenario A, the stellar mass of Procyon A is reproduced within $1\sigma$ uncertainties, with a relative difference of 4\% when $R$ is neglected and 1\% when it is included in the optimization. Scenario B also shows improvement, but the inferred stellar mass still deviates by roughly 10\%.
The $\chi^2$ values from the optimization show a substantial improvement compared to Test 1.
To illustrate the frequency fits, we plotted the $r_{02}$ ratios and the \'echelle diagrams of the individual frequencies for the best-fit model in Figs. \ref{fig:ratios_fit} and \ref{fig:echelle_set_test}. Both scenarios fit the ratios reasonably well, but in the \'echelle diagram, the high-frequency modes of Scenario~A deviate slightly from the observations, as expected due to surface effect. In contrast, Scenario~B does not reflect the expected asymptotic behavior in the diagram, and what is inferred can not be explained by the surface effects.
Overall, Scenario~A better reproduces the stellar parameters. It also better matches the observed ridge pattern, and provides a more reliable basis for inferring the stellar properties.

Before selecting the frequency set, we highlight the frequency glitch visible in Fig. \ref{fig:echelle_set_test}, which appears at high frequencies with Scenario A for $\ell=1$ and with Scenario B for $\ell=0$ and 2. 
To assess its impact on the inference and any resulting bias toward Scenario~A, we performed Test~3, excluding the affected high-frequency modes from the optimization. Figure~\ref{fig:diff_set_Al1_Bl02_red} shows the resulting parameter differences.
In this test, Scenario B reproduces the stellar mass of Procyon A when the radius is included in the optimization process, whereas Scenario A shows a slightly larger discrepancy and recovers the expected mass only within $2\sigma$ uncertainties. 
Nevertheless, Scenario A yields a smaller $\chi^2$ and a smaller relative difference than Scenario B. In the \'echelle diagram (bottom row of Fig. \ref{fig:echelle_set_test}), Scenario A performs worse than in Test 2, but still follows the observed pattern more closely than Scenario B, which does not show any significant changes from the previous test. It is important to note that, compared to Test 1, this test represents a significant improvement, suggesting that the optimization issue may arise from fitting the glitch at higher frequencies, and hence in missing physics in the stellar models, rather than in the possible mixed-mode at low frequencies for $\ell=1$.

Tests 2 and 3 indicated that Scenario A best fits Procyon A, while \cite{White2012} favored Scenario B. However, other modeling studies of Procyon A \citep{Dogan2010,Guenther2014,Compton2019}, reproduced its stellar mass adopting the frequencies of Scenario A.
There are two possible explanations. One is that inaccuracies in the stellar models are compensated for by Scenario~A.
However, since the previous studies \cite{Dogan2010,Guenther2014,Compton2019} and the present one use different input physics in the stellar models, and all favor Scenario A, model inaccuracies alone are unlikely to explain the discrepancy.
This supports the idea that Scenario~A may represent the correct frequency identification.
Another possibility is a misidentification of the observed modes, which could make the inferences for Scenario~B inconsistent and those for Scenario~A more coherent.
This issue remains under debate and we hope that forthcoming observations from the Transiting Exoplanet Survey Satellite (TESS; \citealt{TESS}) will provide a definitive answer.
Although \citet{White2012} found observational evidence favoring Scenario B based on the ridge identification in the \'echelle diagram, our modeling results align with previous theoretical studies \citep[e.g.,][]{Guenther2014} in favoring Scenario A. While the statistical difference in terms of raw $\chi^2$ is moderate, Scenario A consistently yields a dynamical mass in better agreement with the precise interferometric value ($1.478\,\Msun$), particularly when the radius is included as a constraint. In contrast, Scenario B often requires stellar parameters that deviate significantly from the spectroscopic or dynamical constraints to achieve a seismic fit. Therefore, we adopt Scenario A, as the seismic constraint for the remainder of this work. Although Scenario A has the smaller $\chi^2$, the reduce value is similar to the other scenario and is not statistically significant. Nevertheless, Scenario A it offers the most physically consistent  agreement with the global observational parameters
(smaller relative difference and reproduce the expected asymptotic behavior of the frequencies).
%Meanwhile, we adopt Scenario A as the seismic constraints, because our results indicate that this offers the most consistent agreement with the observations \textbf{(}. 
In the following sections, we will adopt the result from test 2 including $R$ as reference, since it seems to provide the best characterization of Procyon~ A. For the surface corrections test, we exclude the $\ell=1$ modes from the optimizations and keep $R$ as a constraint for consistency.

\section{Testing surface corrections}
\label{sec:results}

\begin{table*}[]
\centering
\caption{Inferred properties for the different surface corrections. The $\chi_\mathrm{total}^2 $ and $\chi_\mathrm{freq}^2$ are from the best model provided by AIMS and there respective reduced value.}
%\resizebox{\textwidth}{!}{%
\begin{tabular}{cccccccccccc}
\hline
Test & $M~(\Msun)$ & $L~(\Lsun)$& $R~(\Rsun)$ & $\log(g)$~(dex) & $\chi_\mathrm{total}^2$ & $\chi_\mathrm{freq}^2$ & $\chi_\mathrm{total;r}^2$ & $\chi_\mathrm{freq;r}^2$  \\\hline
 BG14$_1$  & 1.379$\pm$0.025& 6.206$\pm$0.222& 2.040$\pm$0.010 & 3.958$\pm$0.004 & 89& 52 &2.5&1.6\\
BG14$_2$& 1.561$\pm$0.028& 8.172$\pm$0.188& 2.051$\pm$0.010&  4.007$\pm$0.005&  23& 21 &0.6&0.6\\
KJ08  & 1.374$\pm$0.028& 6.225$\pm$0.263& 2.037$\pm$0.011&  3.958$\pm$0.005&  98 & 61 &2.7&1.8\\
 SO15   & 1.504$\pm$0.027& 7.887$\pm$0.158& 2.037$\pm$0.010&  3.997$\pm$0.004&  57 & 56 &1.5&1.7\\
 None   & 1.447$\pm$0.036& 7.145$\pm$0.347& 2.047$\pm$0.011&  3.976$\pm$0.007& 101 & 92 &2.8&2.8\\
 Ratios $(r_{02})$   & 1.496$\pm$0.031& 7.744$\pm$0.296& 2.044$\pm$0.009& 3.992$\pm$0.007&  15 & 14&0.7&0.8\\\hline
\end{tabular}%
%}
\label{tab:Procyon_inf_para}
\end{table*}

% \usepackage{multirow}
% \usepackage{graphicx}
\begin{table*}[]
\centering
\caption{Relative and normalize difference of the stellar properties for the tests of the different surface corrections of frequencies.}
\label{tab:diff_surf}
%\resizebox{\textwidth}{!}{%
\begin{tabular}{ccccccccc}
\hline
\multirow{2}{*}{Test} & \multicolumn{2}{c}{Mass} & \multicolumn{2}{c}{Radius} & \multicolumn{2}{c}{$\log(g)$} & \multicolumn{2}{c}{Luminosaty} \\ \cline{2-9} 
 & $\Delta M/M_r$ & $z$ & $\Delta R/R_r$ & $z$ & $\Delta \log(g)/\log(g)_r$ & $z$ & $\Delta L/L_r$ & $z$ \\ \hline
BG14$_1$ & $-0.07\pm0.02$ & 4.01 & $-0.001\pm0.007$ & 0.29 & $-0.024\pm0.016$ & 22.63 & $-0.12\pm0.03$ & 3.80 \\
BG14$_2$ & $0.06\pm0.02$ & 2.98 & $0.004\pm0.007$ & 0.78 & $-0.012\pm0.016$ & 10.48 & $0.16\pm0.03$ & 5.97 \\
KJ08 & $-0.07\pm0.02$ & 3.71 & $-0.003\pm0.007$ & 0.51 & $-0.024\pm0.016$ & 18.52 & $-0.11\pm0.03$ & 3.13 \\
SO15 & $0.02\pm0.02$ & 0.94 & $-0.002\pm0.007$ & 0.60 & $-0.014\pm0.016$ & 14.61 & $0.11\pm0.02$ & 5.29 \\
None & $-0.02\pm0.03$ & 0.85 & $0.002\pm0.007$ & 0.39 & $-0.020\pm0.016$ & 11.61 & $0.01\pm0.05$ & 0.27 \\
Ratios ($r_{02}$) & $0.01\pm0.02$ & 0.58 & $0.000\pm0.006$ & 0.09 & $-0.016\pm0.016$ & 8.87 & $0.10\pm0.04$ & 2.35 \\ \hline
\end{tabular}%
%}
\end{table*}

\begin{figure}
    \centering
    \includegraphics[width=1\linewidth]{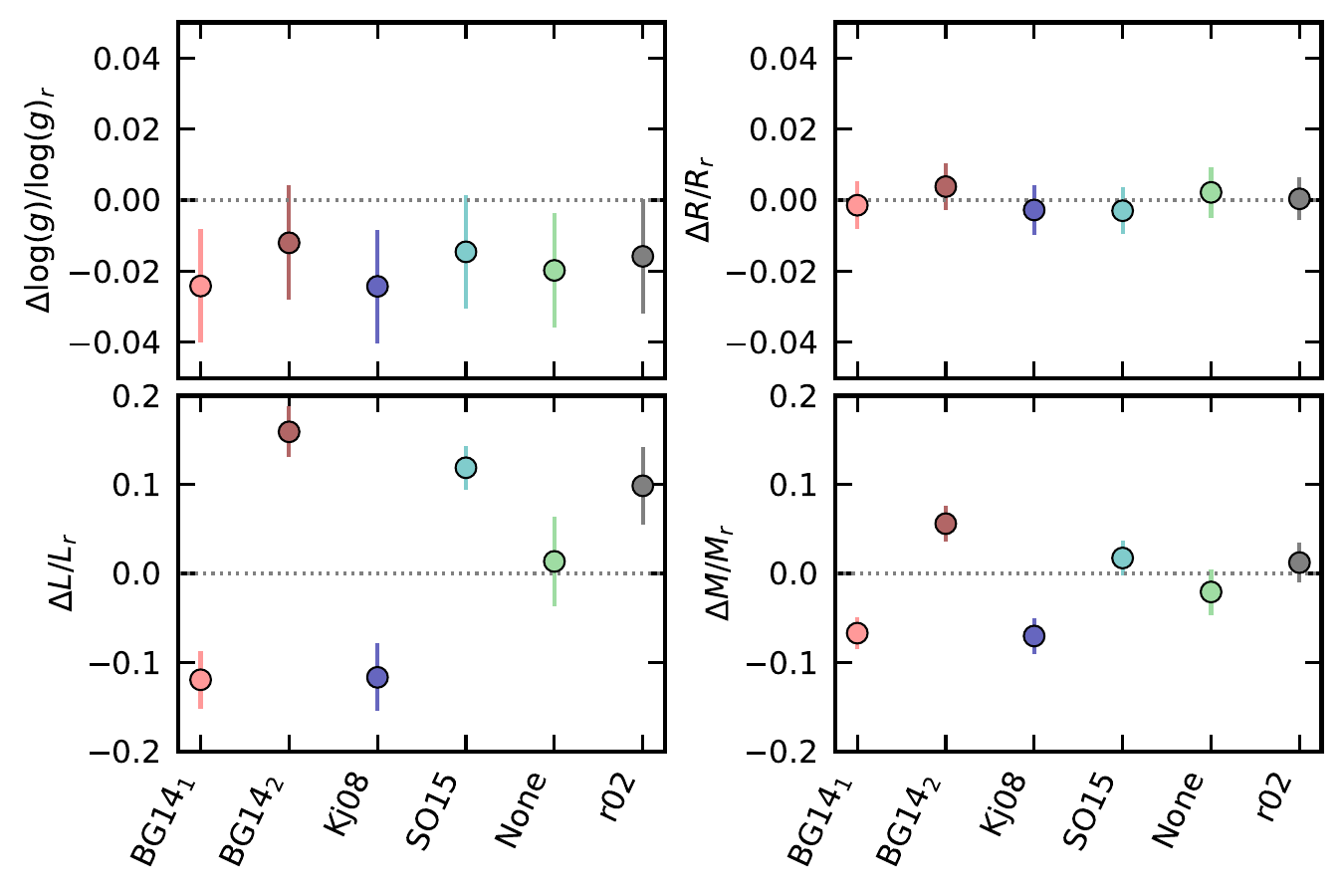}
    \caption{Relative difference for surface gravity (top left panel), radius (top right panel), luminosity (bottom left panel), and mass (bottom right panel) between observation values and our inferred results for the different surface corrections. The dotted line indicate where the relative difference is 0.}
    \label{fig:diff_corrections}
\end{figure}

\begin{figure}
    \centering
    \includegraphics[width=0.9\linewidth]{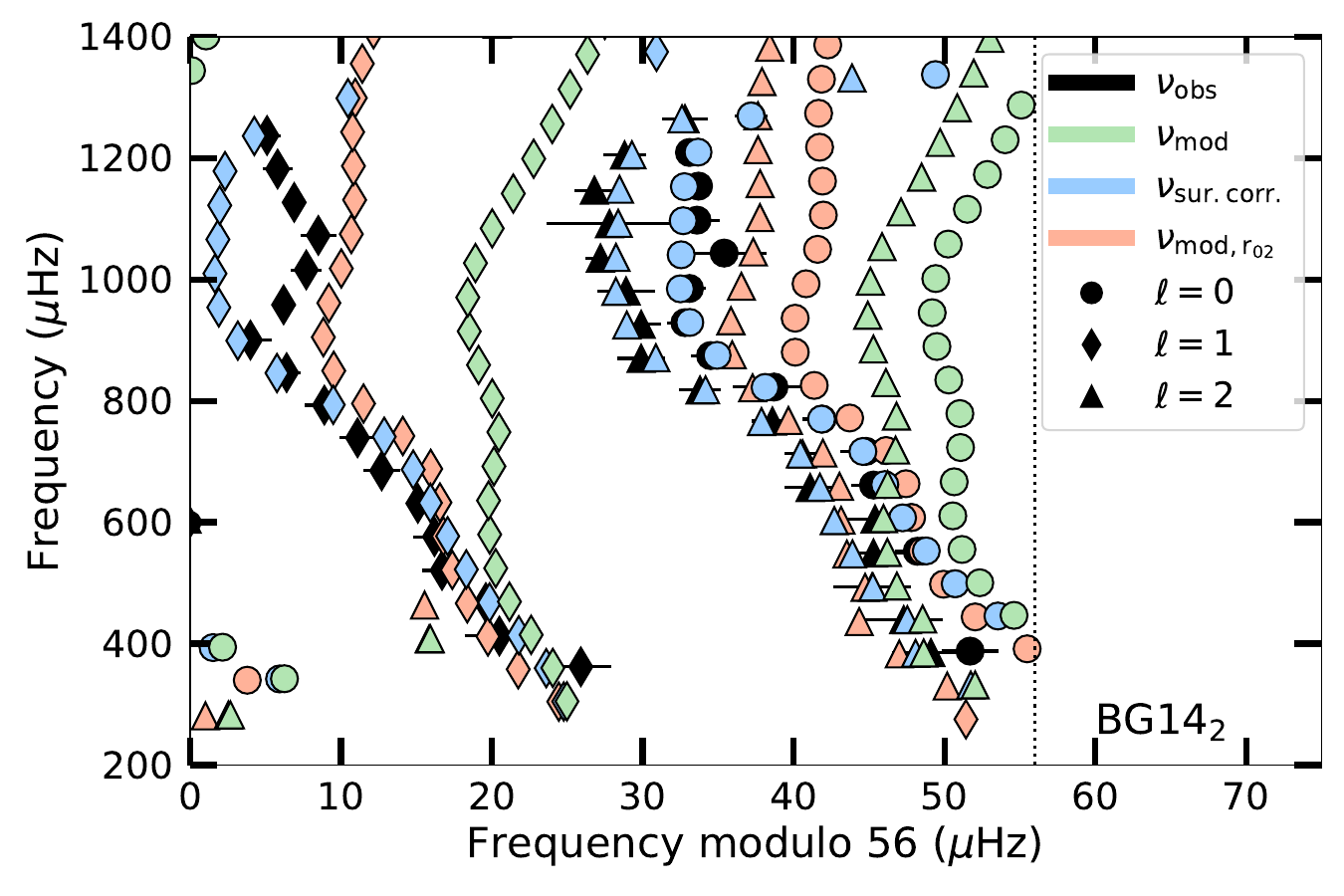}
    \includegraphics[width=0.9\linewidth]{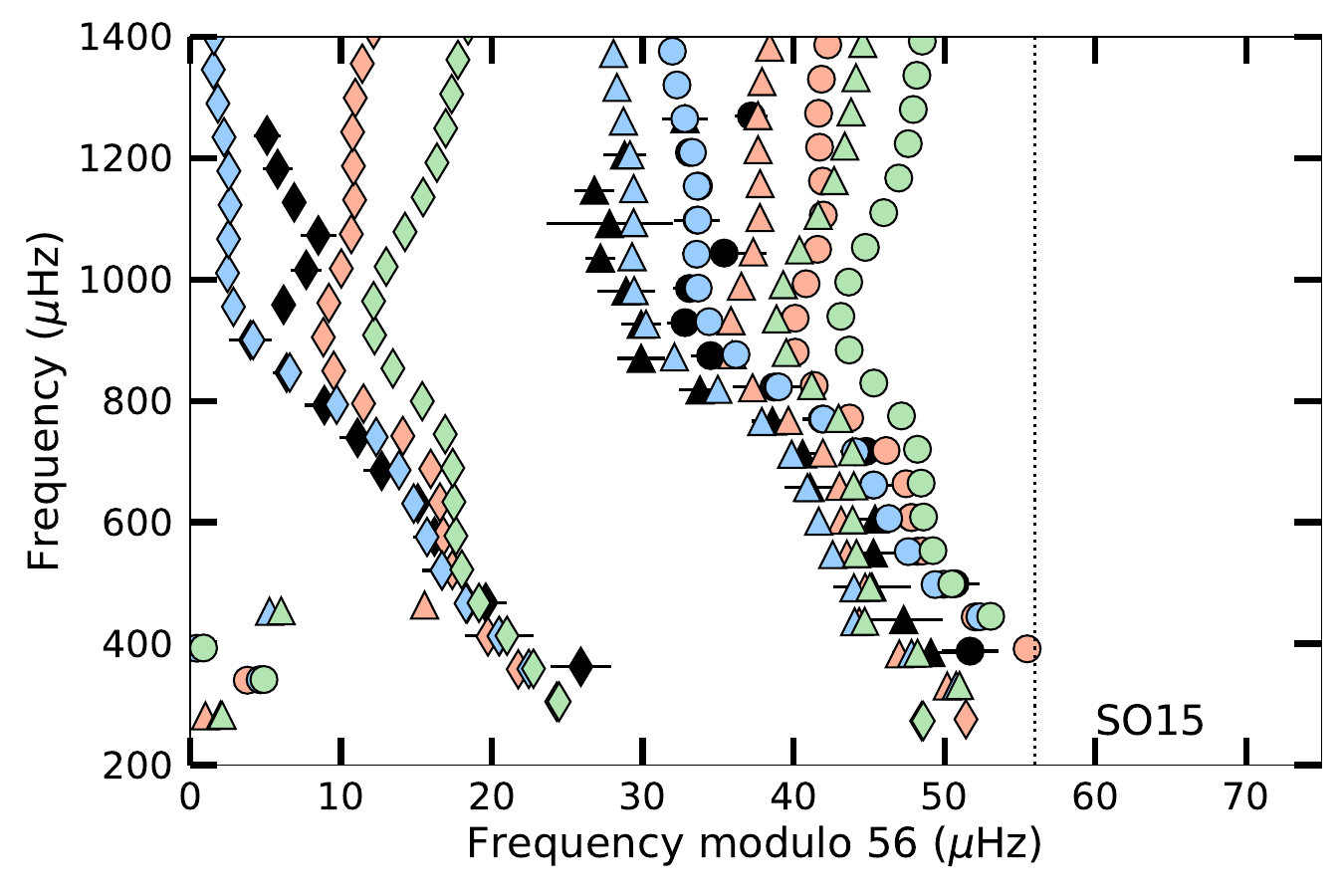}
    \includegraphics[width=0.9\linewidth]{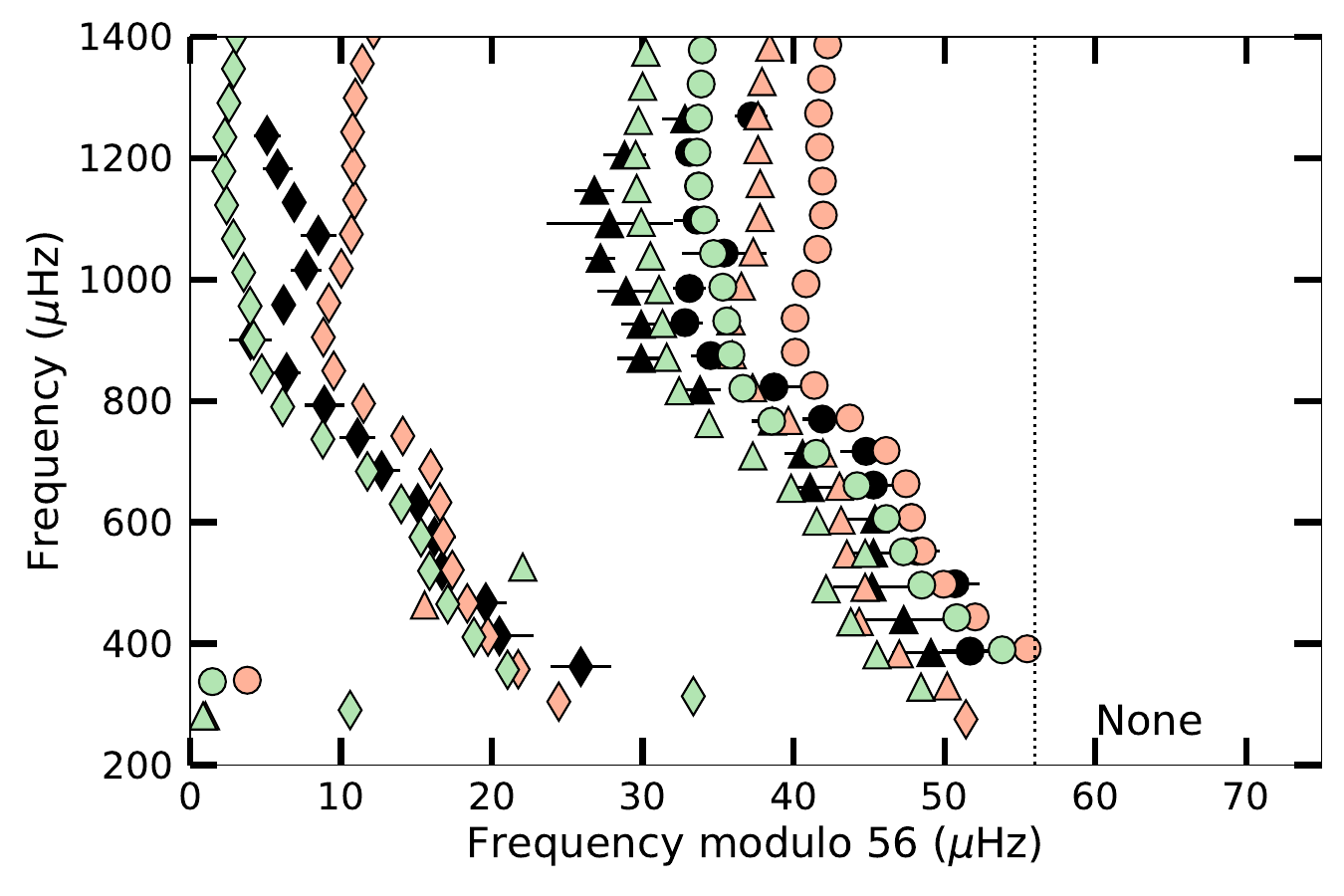}
    \caption{\'Echelle diagram of the frequencies of the models resulting from the optimization procedure. Top panel for the best model with BG14$_2$ correction, middle panel for best-model with SO15 correction and bottom panel for the 
    best-model with no surface corrections. Black symbols with error bars show observed frequencies, green symbols represent the model frequencies without any corrections, blue symbols are the model frequencies with surface correction, and the orange ones are the frequencies of the model constrained by using ratios. The circle represent modes $\ell=0$, triangles represent modes $\ell=1$ and diamond represent modes $\ell=2$. The gray dotted line indicates the 56~$\mu$HZ frequency. }
    \label{fig:echelle}
\end{figure}

In Sec. \ref{sec:senarios}, our tests indicated that Scenario A best reproduces Procyon A. Here, we examine how well the surface corrections under study reproduce the stellar properties of our star. The stellar parameters obtained for each surface correction, along with their respective normalized differences from the observed values, are presented in Tables \ref{tab:Procyon_inf_para} and \ref{tab:diff_surf}.

The tests show that, searching for the best-fit model, the higher $\chi^2_{total}$ values are obtained by fitting frequencies than the ratios. Among the surface corrections, BG14$_1$, KJ08, SO15, show the largest $\chi^2$.
Base on optimization alone BG14$_2$ is preferred,  yielding the smallest  $\chi^2_\mathrm{total}$; neglecting surface corrections yields the worst results. However, comparing the inferred stellar mass to the expected value reveals a 
6\% relative difference with the BG14$_2$, larger than the 2\% deviation obtained with SO15 or with no surface correction. The results with SO15 and with no surface correction reproduce the mass within 1$\sigma$, whereas BG14$_2$ only within 3$\sigma$. 
The other parameters are generally consistent across tests. The radius $R$ is reproduced as expected, since it is used as a constraint. For $\log(g)$, all tests show similar relative differences of 1–2\%. Regarding luminosity $L$, only the no surface correction case matches with a relative difference of 1\%, while all the other corrections show deviations greater than 10\%, with BG14$_2$ performing worst.

The ratios are unaffected by surface effects and therefore can be used as a reference for evaluating the mode identification in each test.
The differences between model frequencies obtained with the ratio-based optimization (orange symbols) and the best models with different surface corrections can be visualized in the \'echelle diagrams in Fig. \ref{fig:echelle}. Not corrected frequencies are indicated by green symbols. In the upper panel, the model corrected with BG14$_2$ (blue symbols) shows the largest deviation compared to the ratio best-model, with frequencies closer to the observed ones. This suggests that the BG14$_2$ surface correction overcorrected the individual frequencies. Although this may produce frequencies that better match the observations, the corresponding stellar properties deviate more from the expected values. 
The smallest deviation is obtained for the model found with the SO15 correction (middle panel), whose inferred stellar mass is closest to the expected value, consistent with the fact that its individual frequencies are closer to the ratio-optimization model. In contrast, the model without surface correction (lower panel) shows frequencies closer to the observed ones than to those of the model optimized with the ratios, pointing to an undercorrection of the surface effects.

Our results indicate that, for Procyon~A, using frequency ratios in the fitting process yields the best results. The ratios provide tighter constraints on the stellar parameters and reproduce values close to the expected ones, particularly for the stellar mass. However, since not all stars have high-quality mode identifications, direct frequency fitting may still be required. Among the surface correction tests, BG14$_2$ fits the observational constraints most effectively but introduces inconsistencies likely related to the surface correction itself, which can shift the frequencies beyond the expected range.
SO15 correction achieves a more accurate recovery of the stellar parameters because, unlike BG14 and KJ08, it is not calibrated on the Sun, but on models covering a broader range of stellar properties. As expected, the case without surface corrections yields the highest $\chi^2_\mathrm{total}$, but still provides reasonable stellar parameters, comparable to those obtained with SO15 or with the ratios. For Procyon~A, we therefore recommend using ratios in the optimization process; if this is not feasible, the SO15 or no surface correction approaches offer the most reliable alternatives.

\section{Conclusion}
\label{sec:conclusion}
%CONCLUDE HERE PLEASE
Procyon A is a benchmark F-type star, with high-quality observations and well-characterized seismic spectra that provide detailed constraints on its internal structure. Its membership in a binary system ensures a well-determined stellar mass, making it particularly valuable for testing and validating stellar models of F-type stars.

Seismic analyses yield two possible scenarios for the individual frequencies: Scenario A and Scenario B. Although this was not the aim of the present manuscript,
using frequency ratios, which are largely insensitive to surface effects, we find Scenario A to be preferred. This scenario reproduces the observed stellar properties, especially the mass, and yields modeled frequencies that closely match observations, consistent with previous studies \citep{Dogan2010,Guenther2014,Compton2019}.
It is important to note that  the use of $\ell=1$ modes  prevents an accurate inference of the stellar properties. This is due to a glitch that appears at higher frequencies. This feature may reflect an incomplete description of the stellar structure, such as core rotation, internal magnetic fields, or an inadequate treatment of core overshoot. 
However, we note that \cite{White2012} preferred Scenario~B as the correct one, in contradiction with our conclusion. 
Certainly,  missing physics in the models or wrong observational parameters identification could have lead to 
prefer Scenario A over B. We wish to point out that to disentangle between the two scenarios,  new analysis and more observations of this star are highly desired.

The frequency ratios provide a useful reference for evaluating the agreement between model and observed frequencies, as they are largely unaffected by surface effects.
This represent the best optimization process, yielding the smallest $\chi^2$ and most accurately reproducing the stellar properties, particularly Procyon A's stellar mass. Testing different surface corrections allows us to minimize discrepancies with the observed constraints.
However, although we could improve the optimization process, this would not improve the inference of stellar properties. Among the surface corrections, the BG14$_2$ achieves the best fit to the observational constraints and the smallest $\chi^2$. However, it cannot accurately infer the stellar mass.
This limitation likely arises because the BG14 correction, developed for the Sun, does not capture the surface physics of F-type stars. The SO15 and no surface correction cases yield higher $\chi^2$ values in the optimization process, but both are able to reproduce the stellar mass of Procyon A.
The improved performance of SO15 likely reflects its calibration over a broad range of stellar models rather than solar ones. The reasonable results from the no surface correction case, despite its high $\chi^2$, suggest that surface effects may be less significant for F-type stars or that other physical processes in the models compensate for the missing surface correction.

The results of the surface corrections can be compared with the missions's core science objectives of
PLATO, which aims to achieve 15\% uncertainty in the mass of solar-type stars. 
All estimated stellar masses fall within this range, with the largest deviations of 7\% with BG14$_1$ and KJ08 corrections. This indicates that BG14$_2$ is acceptable within these uncertainties, as it has a deviation of 6\%. 
However, the deviations observed in some cases, indicate areas where the surface correction models may need refinement to better align with the expected stellar properties.
This study focuses only on Procyon A, so extending the analysis to a larger sample of F-type stars is necessary to fully understand the impact of surface corrections and determine the most appropriate prescriptions. Ultimately, improving our physical modeling of surface effects remains the best path forward. 

Procyon~A is a unique target, as it is one of the few F-type stars with a well-determined stellar mass, providing a strong constraint for stellar models. This makes it a valuable benchmark for improving the modeling of surface layers and for studying convective transport in the outer envelope. Additionally, it offers insights into the stellar interior. The presence of a glitch at high frequencies indicates incomplete modeling of the interior, which could arise from an inaccurate treatment of core overshoot, core rotation, or internal magnetic fields.

\begin{acknowledgements}
We acknowledge funding for the publication of the present manuscript and for the position of N. M. 
from the research theory grant “Synergic tools for characterizing solar-like stars and habitability conditions of exoplanets'' under the INAF national call for Fundamental Research 2023.
\end{acknowledgements}
%-------------------------------------------------------------------------------------------------------------

\bibliographystyle{aa} % style aa.bst
\bibliography{Procyon_paper_vf.bib} % your references Yourfile.bib

%\newpage
%\clearpage
%\onecolumn

\begin{appendix}

\onecolumn

% End longtab 

%\twocolumn
\end{appendix}

%\newpage\phantom{---}
%\newpage

\end{document}